\documentclass[fleqn,twoside]{article}
\usepackage{espcrc2}

\usepackage{graphicx}
\usepackage[figuresright]{rotating}

\newcommand{\AmS}{{\protect\the\textfont2
  A\kern-.1667em\lower.5ex\hbox{M}\kern-.125emS}}

\title{New proposal for numerical simulations of systems with a 
$\theta$-vacuum term\thanks{Talk given by Vicente Azcoiti.}}

\author{Vicente~Azcoiti\address[zaragoza]{Departamento de F\'{\i}sica
        Te\'orica, Universidad de Zaragoza,
        Cl. Pedro Cerbuna 12, E-50009 Zaragoza (Spain)},
        Giuseppe~Di Carlo\address[gs]{INFN, Laboratori Nazionali del Gran
        Sasso, 67010 Assergi, (L'Aquila) (Italy)},
        Angelo~Galante\addressmark[gs]\address{Dipartimento di Fisica 
        dell'Universit\`a di L'Aquila, 67100 L'Aquila (Italy)} and
        Victor~Laliena\addressmark[zaragoza]
}

\begin{document}

\begin{abstract}
A new approach to perform numerical simulations of systems with a 
$\theta$-vacuum term is proposed, tested, and applied to $\textrm{CP}^3$. 
The main new ingredient of this approach is the method used to 
compute the probability distribution function of the topological charge 
at $\theta=0$. We comment also on the possibility of applying this approach 
to $QCD$ at finite baryon density.
\vspace{1pc}
\end{abstract}

\maketitle

The purpose of this contribution is to report a new approach, 
recently proposed by us in \cite{monos}, to simulate 
numerically systems with a $\theta$-vacuum term. 
To fix the ideas let us consider 
the partition function ${\cal Z}_V(\theta)$ of any $\theta$-vacuum like 
model in a finite space-time lattice volume $V$. This action can be 
written, up to a normalization constant, 
as the discrete Fourier transform of the probability distribution 
function (p.d.f.) of the topological charge at $\theta = 0$:
\begin{equation}
{\cal Z}_V(\theta)\;=\;\sum_n\,p_V(n)\,{\mathrm e}^{{\mathrm i}\theta n}\, ,
\label{partition}
\end{equation}
where $p_V(n)$ is the probability of the topological sector $n$. In almost 
all practical cases the sum in (\ref{partition}) 
has a number of terms of order $V$ since the 
maximum value of the topological charge at finite volume is of this order.

The only {\it a priori} reliable 
numerical scheme to analyze the thermodynamics of $\theta$-vacuum like 
models goes through the determination of the p.d.f. 
of the topological charge, $p_V(n)$, and the evaluation of (\ref{partition}). 
But this is a difficult task due 
to the following two technical reasons: 
\textit{i}) 
any numerical determination of $p_V(n)$ suffers 
from statistical fluctuations \cite{PRD}, 
and small errors in $p_V(n)$ can 
induce enormous relative errors in the determination of a quantity as 
${\cal Z}_V(\theta)$ which is an extremely small number of order $e^{-V}$,
\textit{ii}) 
even if we were able to evaluate $p_V(n)$ with infinite precision, the 
sum in (\ref{partition}) contains terms running from 1 
to $e^{-V}$ \cite{RANDOM}.

In few specific cases one can overcome the sign problem \cite{STURN}. 
However 
previous attempts by other groups to simulate $\theta$-vacuum like systems 
\cite{GENER,PRD}, 
were based on the numerical determination of the p.d.f. 
of the topological charge straightforwardly,
by standard simulations, or by 
more sophisticated methods based on the use of multi-binning and re-weighting 
techniques. In all these attempts, 
artificial phase transitions at a $\theta_c$ 
decreasing with the lattice volume were observed for the two-dimensional 
U(1) gauge theory
at strong coupling as well as $\textrm{CP}^N$ models. 
The origin of this artificial
behavior, which follows from a flattening behavior of the free energy for 
$\theta$-values larger than a certain $\theta_c$, 
was analyzed in \cite{PRD,JAPAN}. Both groups 
agreed that the observed behavior was produced by the small 
statistical errors in the determination of the p.d.f. 
of the topological charge, the effect of which became more and more 
relevant as the lattice volume was increased. In Ref.~\cite{JAPAN} it 
was also noticed that by smoothing the p.d.f. flattening disappears. 

Our approach is based on a new method to 
compute the p.d.f. of the topological charge 
and the use of a multi-precision algorithm in order to compute the 
sum in (\ref{partition}) with a precision as high as desired. 
Let us write the 
partition function (\ref{partition}) as a sum over the density of topological 
charge $x_n =n/V$ and 
set $p_V(n)=\exp[-V f_V(x_n)]$, where $f_V(x)$ is a smooth
interpolation of $-1/V \ln p_V(x_n)$:
\begin{equation}
{\cal Z}_V(\theta)\;=\;\sum_{x_n}\,e^{-V f_V(x_n)}\,
{\mathrm e}^{{\mathrm i}\theta V x_n} \, .
\label{pfdiscrete}
\end{equation}
We will assume that CP is realized in the vacuum at $\theta = 0$ since 
otherwise the theory would be ill-defined at $\theta\ne 0$ \cite{NOS}. This 
implies that $\exp[-V f_V(x_n)]$ will approach a delta distribution 
centered at the origin in the infinite volume limit. 

Let us consider the partition function (\ref{pfdiscrete}) 
on the imaginary axis $\theta= -i h$, and let $f(x)$ 
be the infinite volume limit of $f_V(x_n)$. In the infinite volume limit 
the free energy is given by the
saddle point. Assuming that $f(x)$ has first derivative for any $x$ except 
at most in isolated points, we can write the saddle point equation:
\begin{equation}
f'(x) = h.
\label{saddle}
\end{equation}
Our proposal to compute the function $f(x)$ is based in the following 
three steps: 

\begin{itemize}
\item[i.] 
To perform standard numerical simulations of our system at imaginary 
$\theta = -ih$ and to measure the mean value of the density of topological 
charge as a function of $h$ with high accuracy (typically a fraction of 
percent). This is feasible since the system we have to 
simulate has a real action. Then, Eq.~(\ref{saddle}) is used to get a 
numerical evaluation of $f'(x)$.  
\item[ii.] 
To get $f(x)$ we have to integrate $f'(x)$. 
Between the possible 
ways to do this integral, we decided to fit $f'(x)$ 
by the ratio of two polynomials, whose order is chosen to obtain a
high quality fit, and then to perform analytically the integral 
of the fitting function. In this way we get a very precise determination of 
$f(x)$, which allows us to compute the p.d.f. 
in a range varying several thousands 
orders of magnitude. 
This is the main advantage of our approach when compared with other 
methods based on a direct computation of $p_V(n)$.
\item[iii.]
To use a multi-precision algorithm to 
compute the partition function (\ref{pfdiscrete}) using as input the function 
$f(x)$ previously determined.
\end{itemize}

The function $f(x)$ obtained in step ii) suffers from statistical and 
systematic errors, the last coming from the fact that the saddle point 
equation (\ref{saddle}) has finite volume corrections. An analysis of these 
errors for the models and sizes we have studied, 
shows that systematic 
errors due to finite volume effects are smaller than the statistical ones 
in the whole relevant range of $x$. This is the reason why  
we will replace $f_{V}(x_n)$ in equation (\ref{pfdiscrete}) by its asymptotic 
value $f(x_n)$ in what follows. This substitution has no effect in 
the infinite volume limit at imaginary $\theta$ and we are assuming that the 
same holds for real $\theta$.

The x-interval used in the fits is an interesting point since 
it allows to improve our approach. 
From the definition of the partition function it is obvious that we
do not need to know $f_V(x)$ for all the possible values of the topological
charge density $x$: in the large volume limit only the values of $x$ such that
$f_V(x) \le g_V(\pi)$ (where $g_V(\theta)$ is the exact free energy 
density normalized to zero at $\theta=0$) are relevant for the 
computation of the free energy density. 
This allows in practice to fit 
only the relevant part of the data having very good fits with 
a suitably chosen low parameters function.

To test these ideas we have analyzed three models:
the one-dimensional antiferromagnetic 
Ising model within an external imaginary magnetic field, the two-dimensional
compact U(1)  
model with topological charge, and $\textrm{CP}^3$ in two dimensions.

From a numerical point of view the determination of the free energy density 
and order parameter through equation (\ref{partition}) in the Ising  model
has the same level of complexity of more sophisticated models. Furthermore, 
in contrast to 
two-dimensional $U(1)$ gauge theory, where the p.d.f. of 
the topological charge is nearly gaussian, the non-gaussian behavior of 
the p.d.f. of the mean magnetization in the 
antiferromagnetic Ising model makes this model a good 
laboratory to check the reliability of our
approach. We refer the interested reader to \cite{monos} for details of the 
simulations, but would like to notice that 
the p.d.f. of the order 
parameter for such a system takes values in a range of around 2000 orders of 
magnitude. Notwithstanding that, we were able to reproduce the order parameter 
in the whole $\theta$ interval within a few percent \cite{monos}.
The agreement between analytical and numerical results in compact $U(1)$ 
was even better (within a few per thousand).
Furthermore the 
flattening found in \cite{PRD} for the free energy density in relatively 
small lattices was absent in our simulations even in the 80$\times$80 
lattice \cite{monos}.

The last model we have analyzed is $\textrm{CP}^3$. It is the standard 
wisdom that this model shares many 
qualitative features with QCD. 
Even if it has not been analytically solved we 
believe it is worthwhile to compare our results with previous existing 
numerical simulations. 
Also in this model, the previous numerical simulations gave
artificial phase transitions 
with a flattening behavior for the free energy density at a 
$\theta_c$ decreasing with the lattice volume \cite{GENER,PRD}.
Our results for the order parameter at 
$\beta=0.6$ on a $100^2$ lattice \cite{monos} have
no trace of the fictitious phase transition 
found in \cite{PRD}. Furthermore the order 
parameter is clearly different from zero at $\theta=\pi$, hence  
CP is spontaneously broken at this $\beta$ value. An open question is
how CP is realized in the continuum limit \cite{BURK}.

Finally we want to comment on the possibility of applying this approach to 
the analysis of QCD at finite baryon density. Indeed the partition function 
of QCD within an imaginary chemical potential belongs to the general class 
of partition functions described by (\ref{pfdiscrete}), where $x_n$ now stands 
for the density of baryonic charge and $\exp[-V f_V(x_n)]$ 
is the probability distribution 
function of the baryon density at vanishing chemical potential. The 
determination of this p.d.f. would allow us to solve QCD at finite baryon 
density. In this case we are involved with what we could call the inverse 
problem i.e., from numerical simulation of QCD at imaginary chemical potential 
one can measure the mean value of the density of baryonic charge and from it 
we can get the free energy density and the partition function of QCD at 
imaginary chemical potential (left hand side of (\ref{pfdiscrete})). The last 
step of this procedure would be to compute, with the help of the 
multiprecision algorithm, the inverse Fourier transform of (\ref{pfdiscrete}) 
in order to get the p.d.f. of the density of baryonic charge. At the moment 
it is not obvious if this procedure would allow us to see the phase structure 
of QCD at finite chemical potential since, as follows from the discussion on 
the relevant x-interval in the determination of the topological charge as 
a function of $\theta$, the range of density of baryonic charge at which 
the p.d.f. can be measured depends crucially on the maximum value of the 
free energy density at imaginary chemical potential. However the field is 
so interesting that it is worthwhile to be investigated.

Similar ideas to those presented in this contribution have been
proposed and promisingly applied to a random matrix model of
finite density QCD in \cite{NIS}.

This work has been partially supported by an INFN-CICyT collaboration and 
MCYT (Spain), grant FPA2000-1352. 
Victor Laliena has been supported by Ministerio de Ciencia y Tecnolog\'{\i}a 
(Spain) under the Ram\'on y Cajal program.


\begin{thebibliography}{99}
\bibitem{monos}
V. Azcoiti, G. Di Carlo, A. Galante and V. Laliena, 
Phys. Rev. Lett. {\bf 89}, 141601 (2002)
[hep-lat/0203017].
\bibitem{PRD}
J. Plefka and S. Samuel, Phys. Rev. {\bf D56}, 44 (1997).
\bibitem{RANDOM} R. Janik, M. Nowak, G. Papp and I. Zahed, 
Acta Phys. Polon. {\bf 32}, 1297 (2001).
\bibitem{STURN}
W. Bietenholz, A. Pochinsky and U.J. Wiese, 
Phys. Rev. Lett. {\bf 75}, 4524 (1995). 
\bibitem{GENER}
G. Schierholz, Nucl. Phys. (Proc. Suppl.) {\bf B42}, 270 (1995).
\bibitem{JAPAN}
M. Imachi, S. Kanou and H. Yoneyama, Prog. Theor. Phys. {\bf 102}, 
653 (1999).
\bibitem{NOS}
V. Azcoiti and A. Galante, Phys. Rev. Lett. {\bf 83}, 1518 (1999).
\bibitem{BURK}
R. Burkhalter, M. Imachi, Y. Shinno, and H. Yoneyama, 
Prog. Theor. Phys. {\bf 106}, 613 (2001).
\bibitem{NIS}
K.N. Anagnostopoulos and J. Nishimura, hep-th/0108041.
J. Ambjorn, K.N. Anagnostopoulos, J. Nishimura and 
J.J.M. Verbaarshot, hep-lat/0208025.
\end{thebibliography}
\end{document}